\documentclass[11pt]{article}

\usepackage{geometry}
\usepackage{currfile}
\usepackage{textcomp}

\usepackage{xcolor,soul}
\sethlcolor{yellow}

\usepackage[pdftex]{graphicx}
\DeclareGraphicsExtensions{.pdf,.jpeg,.png}

\graphicspath{{./}}

\date{March 19, 2019}

\begin{document}

\title{\LARGE{Decentralized Trusted Computing Base\\ for Blockchain Infrastructure Security}\\
~~}
\author{
\large{Thomas Hardjono$^1$ and Ned Smith$^2$}\\
\large{~~}\\
\large{$^1$MIT Connection Science}\\
\large{Cambridge, MA, USA}\\
\small{\tt hardjono@mit.edu}\\
\large{~~}\\
\large{$^2$Intel Corporation}\\
\large{Hillsboro, OR, USA}\\
\small{\tt ned.smith@intel.com}\\
\large{~~}\\
}

\maketitle

\begin{abstract}
There is a growing interest today in blockchain technology as a possible foundation for
the future global financial ecosystem.
However, in order for this future financial ecosystem to be truly global,
with a high degree of interoperability and stability,
a number challenges need to be addressed related to infrastructure security.
One key aspect concerns the security and robustness
of the systems that participate in the blockchain peer-to-peer networks.
In this paper we discuss the notion of the decentralized trusted computing base
as an extension of the TCB concept in trusted computing.
We explore how a decentralized TCB
can be useful to (i) harden individual nodes and systems in the blockchain infrastructure, 
and (ii) be the basis for secure group-oriented computations 
making within the P2P network of nodes that make-up the blockchain system.
~~\\
\end{abstract}

\newpage
\clearpage



\tableofcontents

\par\noindent\rule{\textwidth}{0.4pt}

\section{Introduction}

There is a growing interest today in blockchain technology as a possible foundation for
the future global financial ecosystem.
Significant media attention has been placed this new field,
and various future visions for the ``blockchain economy'' has been put forward by various authors.
The Bitcoin system proposed by~\cite{Bitcoin} is now nearing 10 years old,
and the BTC currency remains one of the most popular speculative crypto-currencies.

One revolutionary aspect of the Bitcoin system
is its openness for any entity to participate 
in the act of ``mining''.
Any entity can independently and anonymously participate by
computing the ``proof of work'' (PoW) as part of establishing 
consensus regarding the shared ledger among the miners.
As such, the membership in the Bitcoin network is not defined
by geographical location.
This in itself is a radical departure from
traditional enterprise networks consisting of all authenticated entities.
This independence and anonymity of the mining nodes means that it is difficult or even impossible 
to know which nodes participated in a given proof of work computation instance.
This in turn leads to the possibility of anonymous mining pools amassing CPU power (hash power),
which can be unleashed at the opportune moment to skew or influence the network.

If the global financial industry is to make use of blockchain technology for their future infrastructure,
then several deficiencies of the current blockchain systems need to be addressed
before blockchain technology can replace the existing financial infrastructures.
We believe that an important part of these future infrastructures
are the features that are founded on the notion of the {\em decentralized trusted computing base} (DTCB) model.
In this paper we explore how the DTCB model
can be useful to (i) harden individual nodes and systems in the blockchain infrastructure, 
and (ii) be used to enable secure group-oriented computations 
in the blockchain system.

In the next section we briefly review a number of challenges 
facing the nascent area of blockchain technology.
In Section~\ref{sec:TCB-history} we review the history of the notion of the trusted computing base
that emerged in the 1980s from the DoD Orange Book.
We propose a number of desirable features of DTCB in Section~\ref{sec:DesirablePropertiesTCB},
building on the existing industry experience in trusted computing.
Similar to other infrastructure technology that have moved to the cloud,
parts of the blockchain infrastructure may also move to the cloud environment.
We discuss the role of the TCB in the context of 
virtualized and cloud environments in Section~\ref{sec:VirtualizedCloudEnvironments},
paying close attention how the roots of trust can be extended into these virtualized containers.
In Section~\ref{sec:DTCB-usecase} we discuss an important use-case of the DTCB
related to the interoperability and survivability of blockchain infrastructure,
namely the construction of a secure gateway that interconnects blockchain systems.
We close the paper with some future considerations.

\section{Challenges in Blockchain Systems}
\label{sec:challenges}

The emergence of the Bitcoin system has provided the 
first working example of a ``permissionless'' blockchain system 
pertaining to digital currency. 
The term ``permissionless'' and ``trustless'' today is used typically 
to convey the fact that anyone (end-users and mining nodes) 
are free to join or leave the P2P network at any time. 
As such, the ``membership'' of a permissionless blockchain system 
is dynamic and the perimeter is elastic. 

However, there are a number of limitations to the permissionless blockchain design: 
\begin{itemize}

\item	{\em Unknown participation leading to concentration of hash-power}: 
The ability of a node to independently compute the proof-of-work hash 
challenge implies that at any given instance of time there 
is no information regarding the actual number of nodes 
spending CPU cycles on solving the proof-of-work. 
This further implies that permissionless systems 
like Bitcoin can be subject to the amassing of CPU power by unknown entities.
Such entities could conceivably use this concentration of hash-power
to skew or manipulate the network.

\item	{\em The trustless model achieves limited trust}: 
The trustless model based on nodes independently 
computing the proof-of-work achieves only limited trust (technical-trust).
This is because the trustless model has no notion of {\em security quality}
or of strong proofs of security.
The implication of this deficiency is that service agreement or contracts
cannot be established for a blockchain system employing the trustless model.

\item	{\em The value of transactions limited by lack of business trust}: 
The lack of service agreement or contracts means that businesses cannot
count on the available or performance of a given blockchain system.
This is in contrast to the Internet today, which is composed
of a set of ISPs (e.g. local, metro and backbone) which operate
based on peering agreements and SLAs.
Without service agreement or contracts, there is 
a human phycological  barrier limiting the amount of risk 
a business is willing to take (see~\cite{TradecoinRSOS2018}).

\end{itemize}

Currently blockchain systems follow one of the following general 
approaches to the membership of nodes:
\begin{itemize}

\item	{\em Pre-identified participation}: 
Nodes must be identified and authenticated prior to joining the community.  
This does not imply, however, that a node will participate 
in every instance of a consensus computation. 
Examples of these are certain implementations of Hyperledger Fabric (see~\cite{AndroulakiBarger2018}).

\item	{\em Anonymous participation}: 
In this approach the nodes are not identified or 
authenticated as a member of the community.  
Thus, any entity with computing resources can operate a node 
within the community, and the node can ``come and go'' as it pleases.  
This approach is exemplified by the Bitcoin system.

\item	{\em Anonymous-verifiable participation (hybrid)}: 
In this approach a node is able to cryptographically prove it is a member of a blockchain system
without revealing its full identity
(for example, see~\cite{HardjonoSmith2016a} based on the EPID scheme in~\cite{EPID-ISO-20008}).

\end{itemize}

\section{The TCB Model: A short History of Trusted Computing}
\label{sec:TCB-history}

\subsection{Orange Book Trust}

In December of 1985 the U.S. Department of Defense published the 
Trusted Computer System Evaluation Criteria (TCSEC) that 
defined Trusted Computing Base (TCB) as ``the security-relevant portions of a system''. 
All subsequent expressions of trustworthy computing and security policy is described 
in terms of impact and relevance to the TCB. The TCSEC notion of TCB 
is most easily understood if it behaves as a centralized trusted entity. 
Indeed, those systems that achieve the highest-level security 
certification and accreditation have a very centralized system design. 
Security Enhanced Linux (SELinux\footnote{https://github.com/SELinuxProject}) embodies this thinking 
in its design for Linux Security Modules (LSM) 
where security relevant operating system events must 
satisfy LSM imposed security policies. 
This approach is made feasible largely due to the Linux architecture 
that regards everything in the system as either a {\em subject} or an {\em object}. 
Subjects operate on objects and objects expose resources. 
The LSM sits at the confluence of subject-object interaction mandated by 
kernel system calls. 
Using TCSEC conventions, a Linux LSM is the primary TCB component of the operating system.

The TCSEC embraces the notion of peer trusted nodes describing 
them as members of a ...
``Network Trusted Computing Base (NTCB) [which] 
is the totality of protection mechanisms within a network system
-- including hardware, firmware, and software --
the combination of which is responsible for enforcing a security policy [...] 
An NTCB Partition is the totality of mechanisms within a single network subsystem...''. 
The TCSEC NTCB criteria, however, does not include the concept of 
attestation where trustworthiness properties can be automated and dynamically evaluated.

\subsection{The Trusted Computing Group}

The TCSEC criteria focused mostly on operating system security.
However, the operating system is not the sole TCB component in a platform. 
The hardware also plays a significant role as memory page isolation is 
central to the idea of kernel-mode (namely {\em ring-0}) and application-mode (namely {\em ring-3}) security context. 
From the operating system perspective the hardware is 
trusted because it has no alternative way to test and verify that the 
hardware is behaving correctly. 
This does not mean that hardware is not susceptible to compromise,
as indicated by recent exploits such as 
Spectre\footnote{https://spectreattack.com/spectre.pdf} and 
Meltdown\footnote{https://meltdownattack.com/meltdown.pdf}.

The threat of hardware vulnerability motivated the computing industry 
to form the Trusted Computing Group\footnote{https://trustedcomputinggroup.org} (TCG) 
where the notion of a hardware root-of-trust was used to 
distinguish the security relevant portions of a hardware platform. 
The TCG defines trusted computing\footnote{https://trustedcomputinggroup.org/resource-directory/glossary/}  
more organically building upon 
granular components that are described as {\em shielded locations} and {\em protected capabilities}. 
Shielded locations are ``A place (memory, register, etc.) where it is safe 
to operate on sensitive data; data locations that can be accessed only by protected capabilities''. 
Protected capabilities are ``The set of commands with exclusive permission to access shielded locations.'' 

Using these concepts, the TCG breaks down trusted hardware functionality 
into three components, otherwise referred to as ``roots-of-trust''. 
There is a {\em Root-of-trust for Measurement} (RTM) whose primary role is to 
ensure the rest of the platform initializes and boots correctly. 
The {\em Root-of-trust for Storage} (RTS) ensures saved security relevant state, 
cryptography objects (e.g. keys) are persistently available 
for inspection and use regardless of whether the surrounding 
platform firmware and software has been compromised. 
The third root is called {\em Root-of-trust for Reporting} (RTR) 
that contains cryptographic algorithms, protocols and access to 
RTS protected keys so that peer nodes can assess 
trustworthiness properties dynamically. 
This assessment is referred to as {\em attestation}.

\subsection{The Trusted Platform Module}

According to the TCG, a Trusted Platform Module (TPM) is formed 
by combining an RTS and RTR, but the TPM specification does not 
clearly delineate which features are ascribed to RTS vs. RTR. 
Generally speaking, a TPM provides four security capabilities:
\begin{enumerate} 

\item	cryptographic libraries including a true random number generator (TRNG), key generations and key storage; 

\item	measured boot using Platform Configuration Registers (PCR) 
where firmware and software hash values are supplied by a RTM. 
The RTM is supposed to be securely connected to the TPM; 

\item	sealing and binding where data encrypted using TPM keys 
are locked to the platform containing the TPM. 
Sealing expects the PCRs will have a prescribed 
value as an additional constraint upon decryption; 

\item	remote attestation uses TPM keys to sign PCR values 
that are delivered to a remote entity for evaluation. 
The signing key is expected to be derived during 
TPM or platform manufacturing, certified by a certificate authority (CA) 
so that the certificate helps establishing provenance of the physical platform.

\end{enumerate} 

The first successful version was TPMv1.2 that supported a `one-size-fits-all'
approach that primarily targeted the PC market. 
It defined a single storage hierarchy protected using a 
single storage root key (SRK). 
Authorization could be asserted using an HMAC, PCR value, 
locality or assertion of physical presence. 
It supported unstructured Non-Volatile RAM storage and
a variety of cryptographic algorithms including SHA-1, 
RSA, AES, 3DES, MGF1 mask generation and 
Direct Anonymous Attestation (DAA) which is a zero-knowledge algorithm. 
It mandated 32 PCRs that use SHA-1 hash computations.

A second generation TPM v2.0 was not backward compatible with TPMv1.2 
and expanded trusted computing features to better support vertical markets. 
TPMv2.0 introduced platform specific profiles that define mandatory, 
optional and excluded functionality for PC Client, Mobile and Automotive-Thin platform categories. 
Platform-specific profiles allow TPM vendors flexibility in implementing TPM features 
that accommodates a specific market. 
Additionally, TPMv2.0 supports three key hierarchies; storage, platform and endorsement. 
Each hierarchy can support multiple keys and cryptographic algorithms. 
Password-based authorization was added and greater flexibility for 
policy-controlled use of the other authorization mechanisms. 
NV-RAM expanded to support monotonic counters, bitmaps and `extend' operations 
in addition to unstructured data storage. 
Support for stronger cryptographic algorithms was added 
including SHA256 for hashing and RSA, ECC using Barreto-Naehrig 256-bit curve and NIST P-256 curve. 
The 128-bit AES key size became mandatory. 
TPMv2.0 also expanded the number of hash algorithms used to compute PCR values. 
For example, both SHA-1 and SHA256 hash values can 
accompany the same PCR register for added security and interoperability.

\subsection{Intel SGX Root of Trust}

The Intel Software Guard Extensions (SGX) (see~\cite{McKeen2013}) 
offers another perspective on trusted computing base 
where a trusted environment exists within a user process called an {\em Enclave}. 
The SGX TCB consists of hardware isolated memory pages, 
CPU instructions for creating, extending, initializing, entering, exiting and attesting 
the enclave and privileged CPU modes for controlling access to enclave memory. 
SGX takes a layered approach to TCG design where CPU hardware and microcode 
make up the bottom layer consisting of Enclave Page Cache (EPC), 
EPC Map (EPCM) and protected mode operation logic. 
We refer to this as the CPU TCB. 

A second layer TCB consists of SGX runtime code that includes 
a user or ISV supplied SGX runtime. 
We refer to this as the ISV TCB. 
Finally, enclave runtimes may dynamically load code and configuration 
data that further specializes enclave behavior. 
We refer to this as the application APP TCB. 
Intuitively, from the application's point of view, 
application functionality within the enclave is the subset 
that the application developer designates as trusted. 
For example, it may contain sensitive application data, 
algorithms and cryptographic keys off limits to other processes and enclaves. 

The three TCB components, together, make up the trusted subset of 
all processes that is trusted, collectively known as the PLATFORM TCB. 
There is a trust dependency within the PLATFORM TCB, 
as APP TCB functionality must trust ISV TCB functionality 
and ISV TCB must trust CPU TCB functionality. 
However, in the broader context, all SGX TCB elements 
do not need to trust external ring-3, ring-0 and VMX root functionality.

SGX architecture supports helper enclaves known as 
{\em Architectural Enclaves} that perform various 
trusted computing services common to most all application enclaves. 
These service enclaves include: 
the Platform Configuration Enclave (PCE) 
that facilitates provisioning certificates for identity and attestation; 
the Quoting Enclave (QE) for performing enclave-to-enclave attestation 
as well as remote attestation and the Platform Services Enclave (PSE) 
for interacting with IP blocks that exist outside the CPU IP complex. 
Application enclaves may include and exclude architectural enclaves 
as needed to satisfy application objectives. 
In this sense, the PLATFORM TCB is dynamic, 
at least from the application developer's perspective.

A second generation SGX (see~\cite{McKeen2016}) 
added support for dynamic memory management where 
enclave runtimes could dynamically increase or 
decrease the number of enclave pages. 
This is a second form of TCB dynamism where if we consider 
the identity of a TCB to be a function of the 
code or logic executing within the TCB (e.g. cryptographic hash), 
dynamically loaded or unloaded pages also changes the TCB identity. 
Trust decisions that anticipate a particular TCB could become 
confused by dynamic memory management. 
There are at least two ways to address this. 
One approach requires all the pages that could be dynamically 
loaded be included in the hash computation then 
unused pages can be evicted when not needed. 
Alternatively, a TCB identity not tied to a hash function is used. 
For example, the vendor could assign a product name that is 
verifiable using a certificate. 
This approach gives vendors greater flexibility toward code maintenance, 
but at the cost of potentially introducing new vulnerabilities. 
SGX allows both TCB naming approaches but with one important augmentation. 
Each TCB element has a Security Version Number (SVN) 
that the TCB vendor uses to track security relevant changes. 
SVN is a monotonic value designed to detect and, in many cases, 
prevent version regressions. 
It requires a level of discipline on the vendor's behalf to ensure SVN is managed correctly. 
An important benefit of SVN is it reduces the number of unnecessary 
TCB re-certifications due to code maintenance activities having no security consequence.

\section{Desirable Properties of a Decentralized TCB}
\label{sec:DesirablePropertiesTCB}

There are a number of core properties desirable for a 
Decentralized TCB (DTCB) model. 
Individually, each participating node in a DTCB instance must possess 
the fundamental properties of trusted computing, 
and more specifically the core aspects of trustworthiness (technical trust). 
In this context it is useful to revisit some key architectural 
designs of the Trusted Platform Module (TPM) (see~\cite{TPM2003}) established 
by the Trusted Computing Group (TCG)
from the late 1990s  which sought to embody (implement) technical trust in hardware.

\begin{figure}[!t]
\centering
\includegraphics[width=1.0\textwidth, trim={0.0cm 0.0cm 0.0cm 0.0cm}, clip]{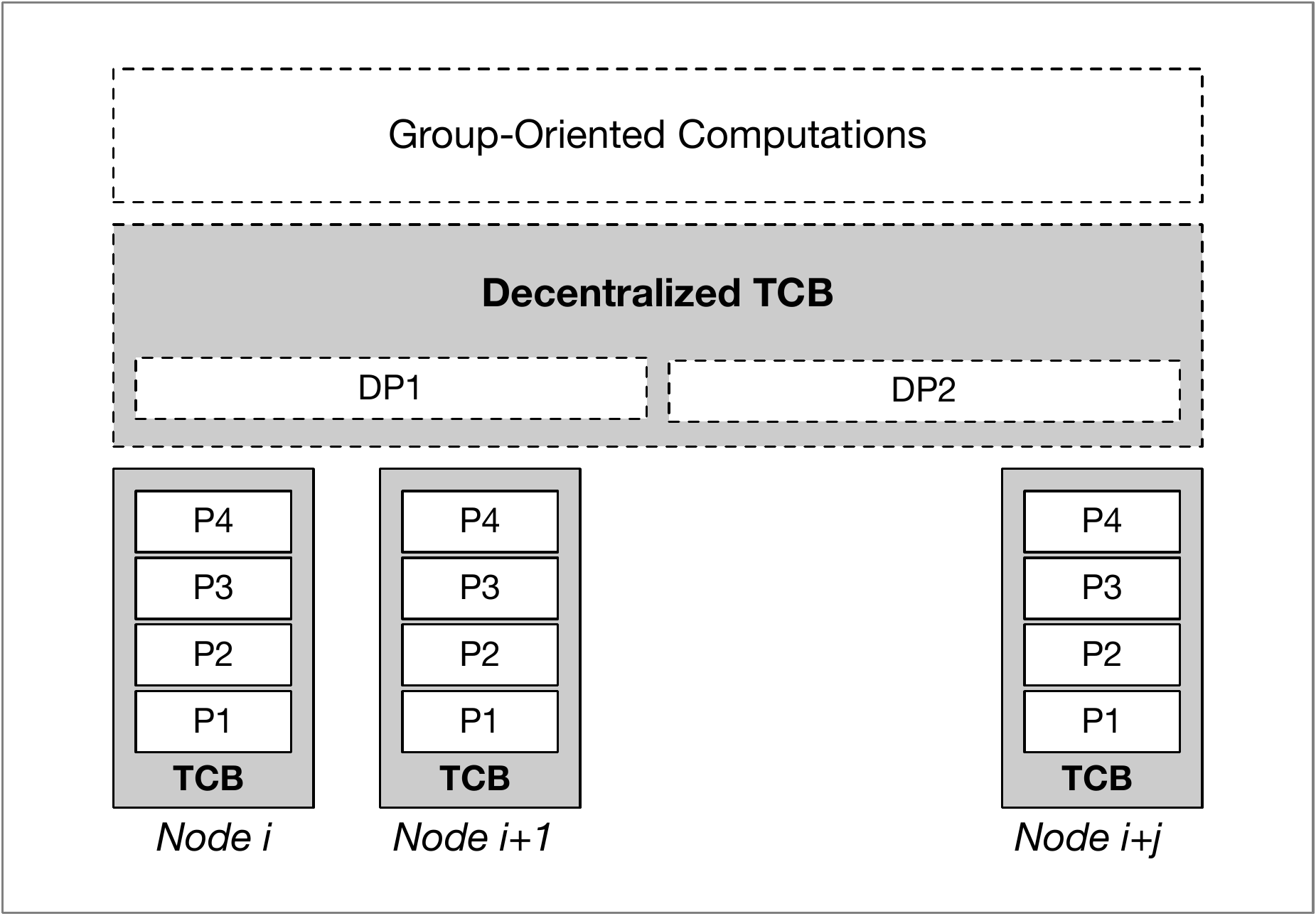}
\caption{Abstract Illustration of Decentralized TCB}
\label{fig:NodesStack}
\end{figure}

\subsection{Properties for Technical Trust}
\label{sec:PropertiesTechnicalTrust}

We say that a node (i.e. system composed of hardware and software)
can be considered to exhibit {\em technical trust} if at least the following properties apply:
\begin{itemize}

\item	{\em TCB Property P1: Performs a well-defined function}. 
The core idea here is that the function being executed by the 
TCB must not harm the TCB itself and must be computationally 
bounded so that it does not consume all available resources.  
In the TPM1.2 design, these functions consisted of a set of 
primitive operations and some cryptographic functions. 
In recent systems such as SGX, there is more freedom 
for the user to load arbitrary functions into the 
trusted execution environment to be executed, 
but the property remains true, 
namely that a user-defined routine must execute 
within the memory space determined by the SGX systems itself.

\item	{\em TCB Property P2:  Operates unhindered and shielded from external interference}.    
In order for an implementation of a function to be useful in a TCB, 
the function must be able to execute until its completion without 
being hindered in any way (e.g. resources locked or made unavailable) 
or that its operations are not skewed or influenced in any fashion.

\item	{\em TCB Property P3: Cryptographic Identity}. 
TCB instances must be distinguishable from each other. 
Trust associations between application layers and 
TCB nodes will differ in deployment; 
unless mirrored for fault-tolerance. 
Cryptographic identity ensures each TCB node 
possesses a unique identity and can prove that identity to a peer node.

\item	{\em TCB Property P4:  Trustworthy TCB Dynamism}.
For a TCB to be practical, it must be able to be {\em updated, expand and contract}. 
Few instances of code are 100 percent error free. 
Those that are proven to be error free are highly constrained 
making them nearly impractical for interesting use cases. 
The most relevant use cases involve establishment of 
a cryptographic identity and firmware or microcode update. 
Beyond these use cases the TCB needs to be dynamic 
allowing layers of functionality to be added (or removed) 
to account for changing hardware and workload computational requirements.

\end{itemize}

We can further say that a node effectively participates 
in forming a DTCB if at least the following properties apply (see Figure~\ref{fig:NodesStack}): 
\begin{itemize}

\item	{\em DTCB Property DP1: Group Membership}. 
A DTCB consists of a group of TCB nodes where 
group membership criteria are applied. 
Membership enforcement ensures a non-TCB node does not 
become a member and compromised or non-compliant 
TCB nodes are expelled from the group. 
For DTCB establishment, the group of TCB nodes must 
cooperatively verify potential DTCB member nodes 
are authorized as members.

\item	{\em DTCB Property DP2: Truthful Attestation}. 
For a DTCB to be practical, it must be able to truthfully report 
the result of its static and dynamic composition, 
internal execution or function and status of its resources 
(e.g. registers, memory usage, etc.).  
Although the industry has defined several attestation solutions 
(e.g. FIDO key attestation, see~\cite{Jones2015};
Android key attestation, see~\cite{Android-key-attestation};
Microsoft TPM key attestation, see~\cite{Microsoft-TPM-2017};
Intel SGX, see~\cite{AnatiGueron2013-SGX}),
they do not consider its role in establishing a DTCB. 
For DTCB establishment, the group of TCB nodes must 
cooperatively verify potential DTCB member nodes' 
TCB trust properties are aligned with vetted DTCB policies.

\end{itemize}

\subsection{Possible Group-Oriented Features}

Based on the above three technical-trust properties, 
there are a number of possible features that maybe achieved using the above properties:
\begin{itemize}

\item	{\em Anonymous Group Membership}: 
Following from property DP1 above, 
the use of hardware to implement the three technical-trust 
properties allows a node to use (reveal) different 
degrees of device-identification and ownership-identification, 
depending on the type of blockchain network it is participating in. 
For example, in the case of permissionless networks where the 
anonymity of nodes is desirable, 
a node may use verifiable 
anonymous identities (see~\cite{HardjonoSmith2016a} and~\cite{Camenisch2002}) to prove it 
is a legitimate member of the network without revealing which member it is.  

In strongly permissioned blockchain systems, 
a node could reveal all manufacturing details of 
the hardware, firwares and softwares to an authorized querier in order to prove 
that the node complies to the minimal operational requirement 
of the permissioned blockchain infrastructure.

\item	{\em Group Reporting}: 
Following from property DP1 above, a group of DTCB nodes
can employ a {\em group-oriented quote protocol} as a counterpart of 
the single hardware quote protocol (see~\cite{TPM2003Design}).
Among others, the group-quote protocol could require each participating node
to perform some internal secret computation,
and report its result using the traditional quote protocol to other members of the group.
A group-shared secret key could additionally
be deployed to protect the transmission of each quote result.

Another more general use of group-reporting
is for a group of nodes to report (to each other)
their respective {\em system manifests}, 
namely a cryptographically signed list of its hardware, firmware and softwares installed 
(for example, see the TCG signed manifest in~\cite{TCG-IWG-2006-Thomas-Ned}). 
This is crucial in cases where nodes are operating different version of 
hardware and software (e.g. version of mining software), 
and where a given version of a mining software may be known to possess
bugs and/or susceptible to malware attacks which may dramatically impact
the blockchain as a whole.

\item	{\em Group Computation Participation}: 
The ability for a node to truthfully report (prove) its 
internal hardware status (e.g. registers) allows the nodes to prove 
that it actively participated a given group-consensus computation. 

New consensus computation algorithms could be designed 
which would embed anonymous-verifiable identification of 
a given node as a precondition for the successful outcome of 
that node's proof-of-work (PoW). 
Similarly, a confirmation broadcasted by a node could require that the node
also anonymously attach proof of its DTCB-capabilities.
\end{itemize}

In the next section we discuss the use of DTCBs in virtualized cloud environments,
motivated by the fact that parts of the blockchain infrastructure (e.g. mining nodes)
may operate in a hosted environment -- which faces a number of security challenges in itself.

\section{Hardware Rooted TCBs in Virtualized Cloud Environments}
\label{sec:VirtualizedCloudEnvironments}

This section describes common hosting environments for {\em containers}, 
another form of distributed system popular among cloud and 
network edge service provider networks. 
It further illustrates techniques for dynamic establishment of 
TCB cryptographic identities and TCB layering. 
When correctly applied, container environments 
can be formed with a hardware rooted TCB otherwise 
known as a {\em hardware root of trust} (RoT). 
Hardware RoT is an essential ingredient in the formation of a decentralized TCB.

\begin{figure}[t]
\centering
\includegraphics[width=1.0\textwidth, trim={0.0cm 0.0cm 0.0cm 0.0cm}, clip]{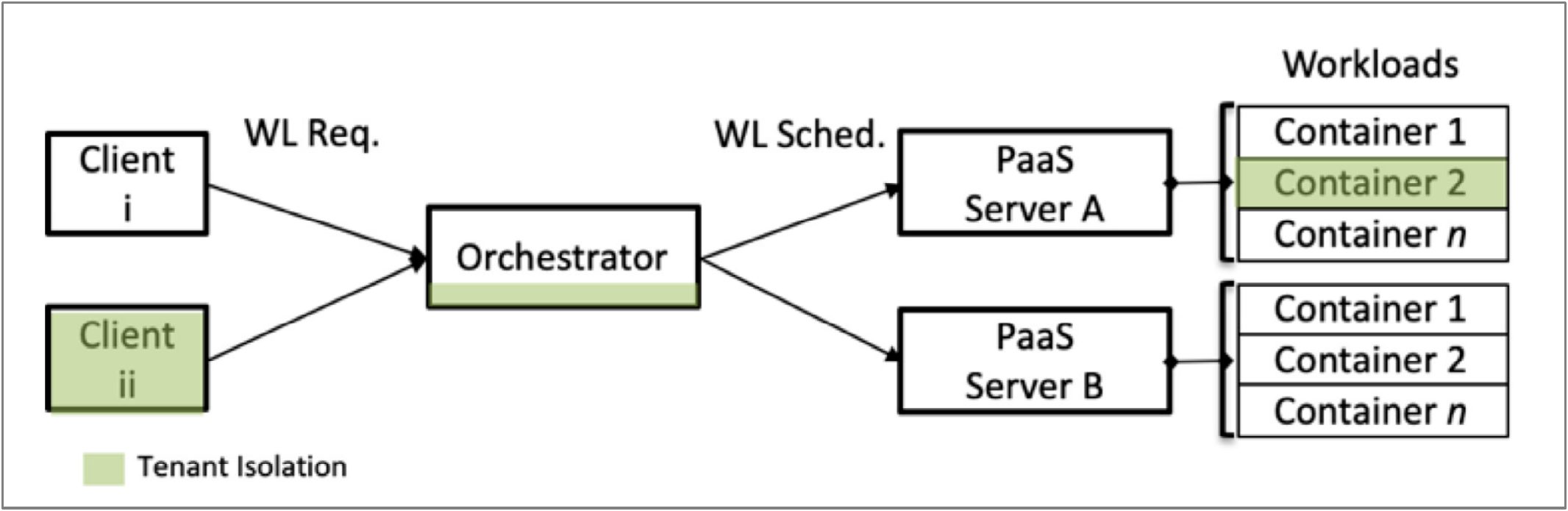}
\caption{Centralized Orchestration of Container Pools}
\label{fig:CentralizedOrchestration}
\end{figure}

\begin{figure}[t]
\centering
\includegraphics[width=1.0\textwidth, trim={0.0cm 0.0cm 0.0cm 0.0cm}, clip]{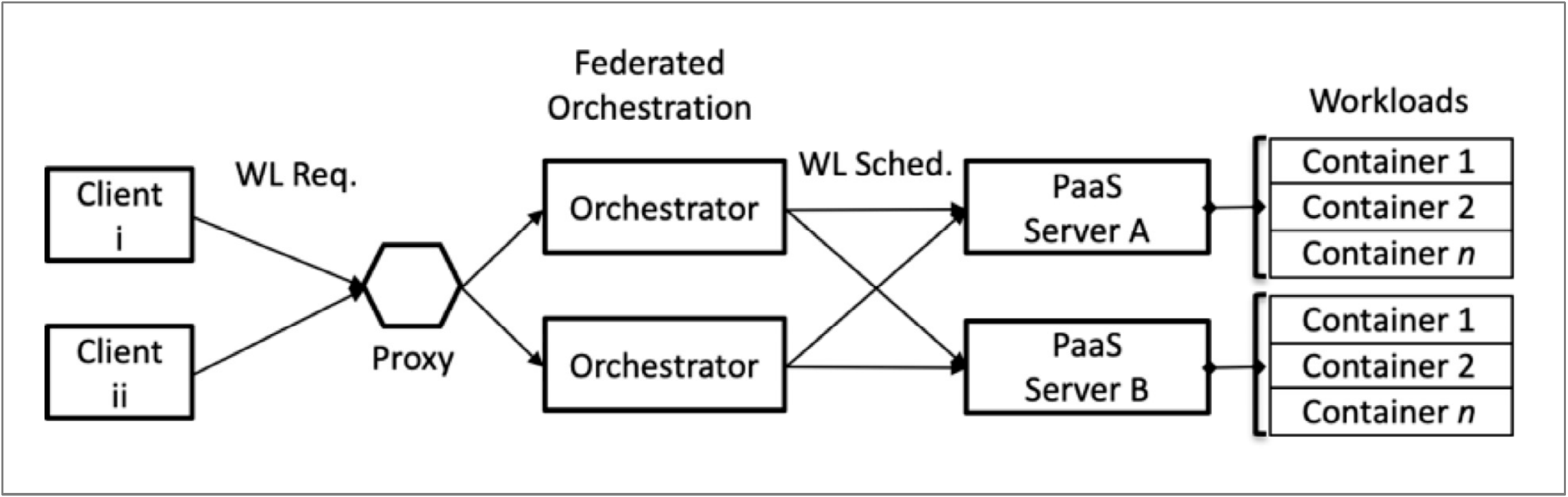}
\caption{Federated Orchestration of Container Pools}
\label{fig:FederatedOrchestration}
\end{figure}

\begin{figure}[t]
\centering
\includegraphics[width=1.0\textwidth, trim={0.0cm 0.0cm 0.0cm 0.0cm}, clip]{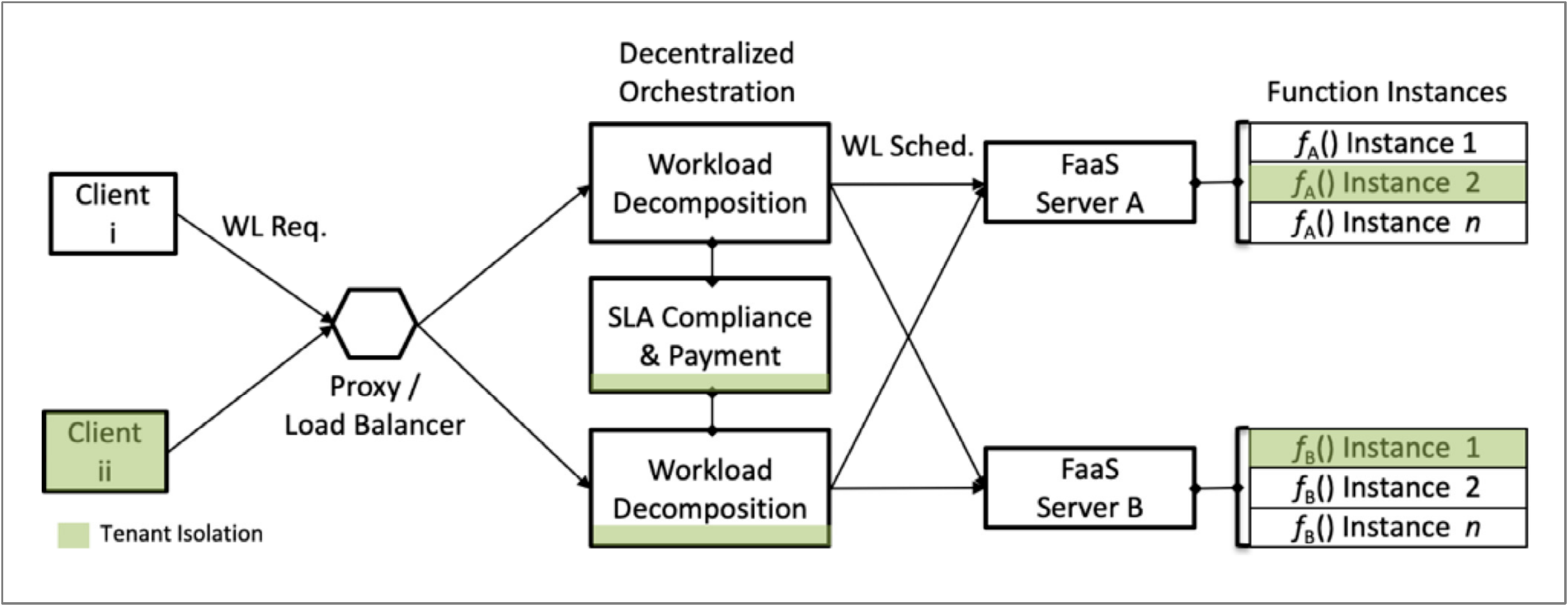}
\caption{Decentralized Orchestration with Decentralized FaaS}
\label{fig:DecentralizedOrchestration}
\end{figure}

A few typical deployment models consist of 
decentralized compute (aka containers) with centralized orchestration (Figure~\ref{fig:CentralizedOrchestration}). 
A more sophisticated model has decentralized, 
but federated orchestration with decentralized pools of containers (Figure~\ref{fig:FederatedOrchestration}). 
Still a third model employs Function-as-a-Service (FaaS) 
to decompose workloads into their functional components, 
where function execution is distributed across multiple 
function provider nodes. Application execution and workflow logic 
encapsulates ``orchestration'' resulting in decentralized 
orchestration with decentralized function execution (Figure~\ref{fig:DecentralizedOrchestration}).

Multi-tenancy is a security challenge facing all deployment models. 
Client workloads are presumed to be mutually suspicious 
and therefore at risk for digital espionage from among the tenant community. 
PaaS, FaaS and orchestration hosting environments are expected 
to provide appropriate tenant isolation to counter potential attacks. 

Additionally, network-based attackers could pose man-in-the-middle 
threats that require end-to-end confidentiality and integrity 
protection of workload payloads as they propagate among the various stations. 
Multi-tenant isolation and end-to-end cryptography technologies 
must be effectively used to ensure comprehensive security. 
Although these security capabilities alone are not enough to ensure reliable operations.

PaaS servers may advertise container pools having 
equivalent workload hosting capabilities, but they may indeed differ. 
Proof of algorithm equivalency is a desirable property when 
interchangeable function and workload hosting environments exist. 
Ideally, a scheduling routine should check function equivalence 
as a precondition of committing the execution resource.

Federated orchestration may require distributed consensus at the orchestration level. 
Workload scheduling involves management of task queues 
where tasks may complete early, late or on time. 
Delivering task results incurs latency from server to client and 
may require temporary caching due to availability of both. 
Task statistics inform regarding resource utilization and performance optimization. 
They are also essential to SLA compliance, accounting and billing processing. 
Decentralized ledger technology can be used to track task scheduling 
statistics so that servers accurately report resource utilizations 
countering potential accounting and billing fraud. 
Federated orchestrators can more conveniently avoid workload server 
oversubscriptions using blockchain to track task queue status.

Decentralized orchestration using Service Level Agreements (SLA) 
that describe compensation for performing the algorithm 
can prove all parties stake is resolved equitably according to the SLA contract. 
In a truly decentralized FaaS system, workload decomposition 
and workload scheduling are themselves functions to be processed by the FaaS fabric. 
SLA compliance checking could involve collection, 
processing and verification of multiple auditing and accounting 
logs spread across dozens or even thousands of compute nodes in a FaaS fabric. 
Any one system failing to keep accurate logs 
potentially disrupts processing across the entire FaaS system. 
Blockchain proof of stake algorithms ensures audit and accounting logging 
is performed with redundancy and guards against isolated cheating. 
SLAs describing compensation for performing the task can prove 
all parties' stake is resolved equitably according to the SLA contracts.

Multi-tenancy, task isolation and end-to-end cryptographically protected workloads, 
SLAs, accounting and telemetry data motivate application 
of trusted computing principles across Edge and Cloud computing infrastructures. 
The cryptographic identities of the various nodes and their roles 
need to resist impersonation attack. 
Cryptographic keys that protect the confidentiality and integrity of 
the data need to resist attacks while in flight and while cached or stored. 
Function executions need to resist attacks on code while it executes. 
To achieve these goals, it is essential that the nodes involved in 
the decentralized Edge and Cloud computing enterprise assess 
trustworthiness properties before placing data, code and resources at risk. 
Attestation procedures applied when computing resources are 
placed into operation and periodically during use ensures expected 
operational integrity characteristics are in place as a 
pre-requisite to decentralized application executions. 
The DTCB is primarily tasked with establishment of trustworthiness pre-requisites.

\subsection{TCB Layering}

DTCB trustworthiness can be understood in terms of its 
roots-of-trust components and its methodology for TCB layering and update. 
The Trusted Computing Group has defined 
Device Identity Composition Engine (DICE) -- see~\cite{TCG2016a} --
which is a trusted hardware building block for 
generating cryptographic device identities and 
attestation using the identities. 
The hardware implementing DICE is the first layer of 
a layered TCB architecture. 
Subsequent TCB layers can be dynamically added or removed 
to fallback to a trusted state. TCB layers may be 
added during manufacturing and later at or during deployment.

The Layered TCB (LTCB) approach seeks to identify 
the most essential trusted computing components 
implementable in hardware and whose implementation is verifiably correct. 
Techniques for dynamic TCB layering are also a consideration for LTCB design.

The following are some considerations for TCB layering:
\begin{itemize}
\item	{\em Hardware Root of Trust (ROT)}: 
The base layer capabilities are trusted and implemented in hardware. 
That is to say they are immutable or that mutability is highly constrained. 
For example, programmable integrated fuses [REF-vi]  
may be set during manufacturing but remain immutable subsequently. 
Algorithms for computing cryptographic one-way functions, 
key derivation and key generation functions can 
have immutable hardware implementations. 
Circuit power-on and bootstrapping control logic can be immutable. 
Other TCB logic may be mutable but only under well-defined conditions. 
For example, a CPU micro-code may be patched post manufacture. 

\item	{\em TCB Layer Identity}: 
Subsequent layered TCB environment is unambiguously distinguishable. 
For example, the product ID of an Intel Core processor with 
virtualization identifies an environment where 
a CPU mode switch causes a hypervisor to execute in VMX Root. 
The microcode patch level or SVN further distinguishes the TCB layer. 
Many Intel CPUs have a monotonically increasing 
Security Version Number (SVN) that changes whenever 
a security relevant change is made to micro-code. 
The loaded hypervisor image also distinguishes the TCB layer. 
Collectively these attributes identify a hypervisor-based TCB. 
However, it does not disambiguate instances of the same TCB. 
It may be necessary to establish a cryptographic identity 
for a TCB layer for on-boarding, resource management, auditing, accounting and telemetry.

\item	{\em Inspect-ability of Next Layer Software}: 
Because TCB identity can be a function of the software that is 
dynamically loaded, the current layer TCB must be 
able to inspect the next layer software. 
Inspection may simply be to compute a hash value for computation of 
a TCB identity or may involve more rigorous proofs of integrity and expected behavior.

\item	{\em Layer Sequencing}: 
Depending on the design of the hardware ROT, layering sequence may be relevant. 
For example, the current layer TCB may exclusively depend 
on the most recent layer for all of its trusted capabilities. 
Other architectures may allow subsequent TCB layers direct access to the hardware ROT.

\item	{\em Layer Attestation}: 
TCB layers may provide security functionality that is 
unique to a platform or system. 
These interactions patterns do not necessarily follow layer sequencing patters. 
Therefore, it may be appropriate to precede inter-layer 
interactions with layer attestations to establish 
a layer's trustworthiness profile. 
Layer attestation primitives also support construction of DTCB as will be discussed later.
\end{itemize}

\subsection{Examples of TCB Layering}

\begin{figure}[t]
\centering
\includegraphics[width=1.0\textwidth, trim={0.0cm 0.0cm 0.0cm 0.0cm}, clip]{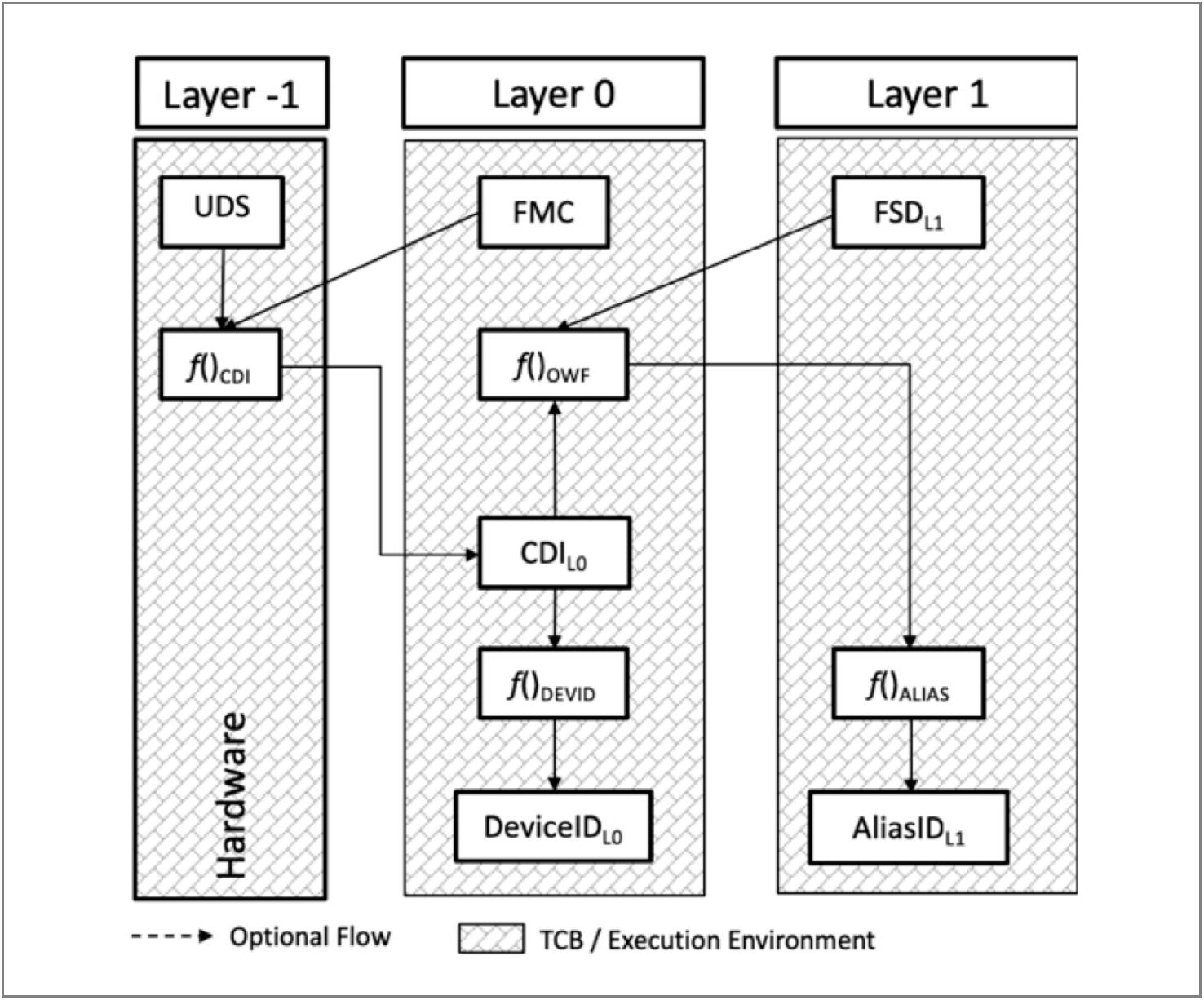}
\caption{Trusted Computing Group -- Device Identity Composition Engine (DICE) layered TCB architecture}
\label{fig:TCG-DICE-Layering}
\end{figure}

This section discusses several examples of TCB layering architectures. 
The first (Figure~\ref{fig:TCG-DICE-Layering}) highlights a DICE architecture consisting of 
hardware (Layer -1) containing two trusted capabilities:
(i) the {\em Unique Device Secret} (UDS) 
and (ii) {\em Compound Device Identifier} (CDI) function. 

The Unique Device Secret (UDS) is a one-time programmable globally unique value. 
Its only use is to seed a Compound Device Identifier (CDI) function that, 
combined with an {\em First Mutable Code} (FMC) value, 
generates a symmetric secret that is specific to the layer that provided the FMC. 
The FMC combined with Layer 0 product ID information identifies the Layer 0 TCB. 
The CDI function is a one-way function  that uses the UDS to produce 
a keyed hash of the FMC called the CDI. 
The CDI {\em uniquely} identifies the Layer 0 TCB.

The CDI is securely installed into the Layer 0 environment 
where it serves two purposes: (i) to seed a one-way function for creating a Layer 1 symmetric secret, 
and (ii) to seed a device identity generation function. 
For example, 
${f()}_{DEVID}$
could be an RSA key generation function 
where CDI is used to seed its random number generator. 
The 
${DeviceID}_{L0}$
is an asymmetric unique Layer 0 identifier 
that may be suitable for a variety of user defined deployment usages.

The Layer 1 TCB is identified using the 
{\em Firmware Security Descriptor} (FSD) which is the firmware hash component of 
the one-way function (i.e. ${f()}_{OWF}$)
found in the Layer 0 TCB 
that computes the Layer 1 symmetric secret used to seed 
the 
${f()}_{ALIAS}$
function that generates 
${AliasID}_{L1}$ that {\em uniquely} identifies the Layer 1 TCB.

\begin{figure}[t]
\centering
\includegraphics[width=1.0\textwidth, trim={0.0cm 0.0cm 0.0cm 0.0cm}, clip]{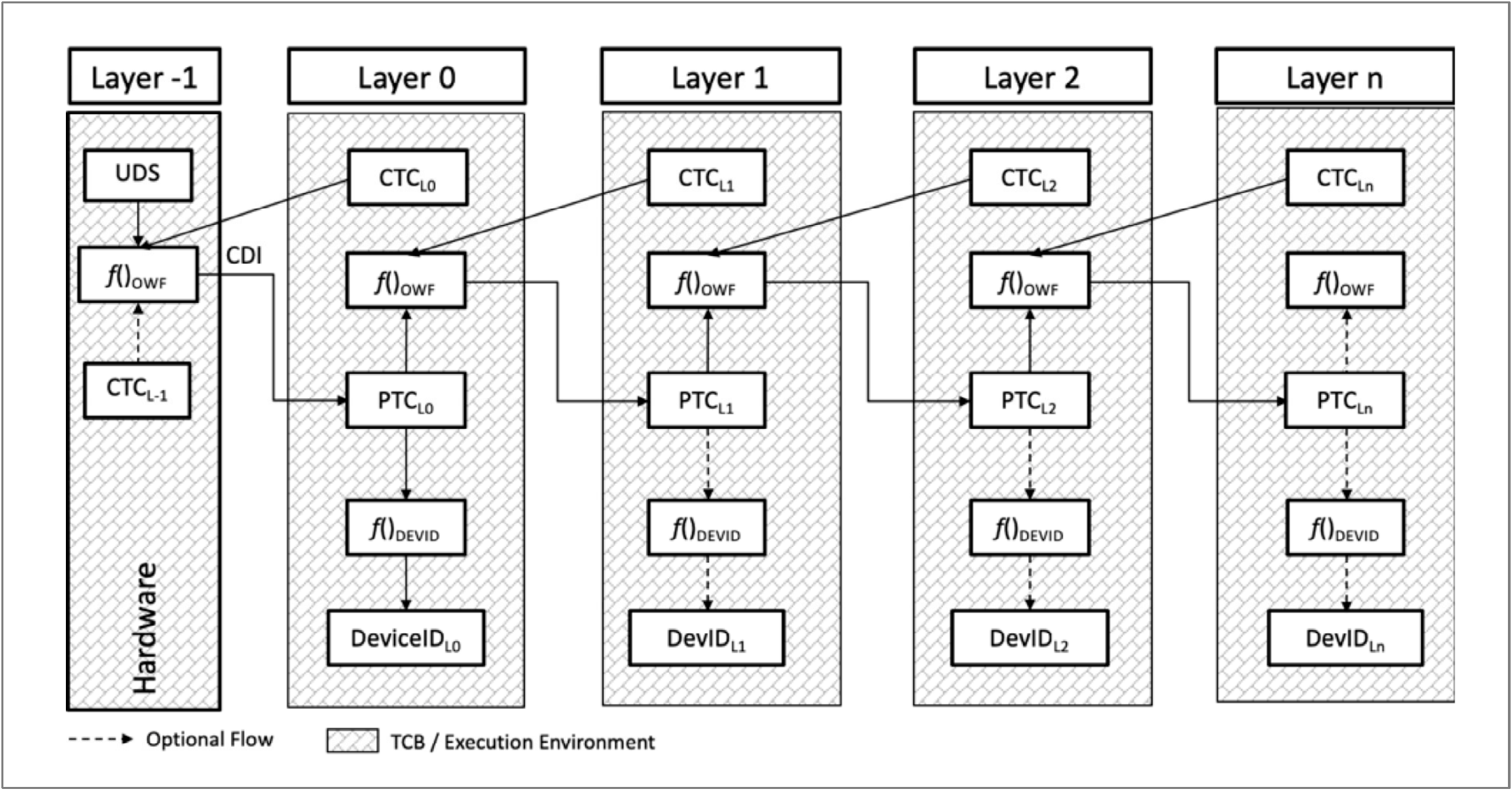}
\caption{A generalized Device Identity Composition Engine (DICE) layering architecture.}
\label{fig:Generalized-DICE-Layering}
\end{figure}

A generalization of the DICE architecture layering (Figure~\ref{fig:Generalized-DICE-Layering}) 
can be inferred following the naming convention where 
the Layer 0 TCB identity is known as the 
{\em Current TCB Context} (CTC) and the Layer 0 unique TCB identity is known as 
the {\em Previous TCB Context} (PTC) 
because it captures Layer -1 TCB layering dependency. 
Although the hardware (Layer -1) TCB does not have 
a previous layer dependency, the UDS provides uniqueness. 
Optionally, the one-way function could accept a Layer -1 TCB 
identity value ${CTC}_{L-1}$ (though not described in [REF-DICE]).

Subsequent layers each rely on its respective previous 
TCB layer to provide a one-way function that 
inspects the current (to be instantiated) layer ${CTC}_{L_{n} }$ and 
the unique 
${PTC}_{L_{n-1} }$
identifier to produce the current layer's unique identifier ${PTC}_{L_{n} }$. 
The PTC value propagates both the platform uniqueness property 
(inherited from the UDS), 
layer uniqueness (UDS + CTC) and layer sequence property; 
which is the combination of all prior ${f()}_{OWF}$ functions.

\begin{figure}[t]
\centering
\includegraphics[width=1.0\textwidth, trim={0.0cm 0.0cm 0.0cm 0.0cm}, clip]{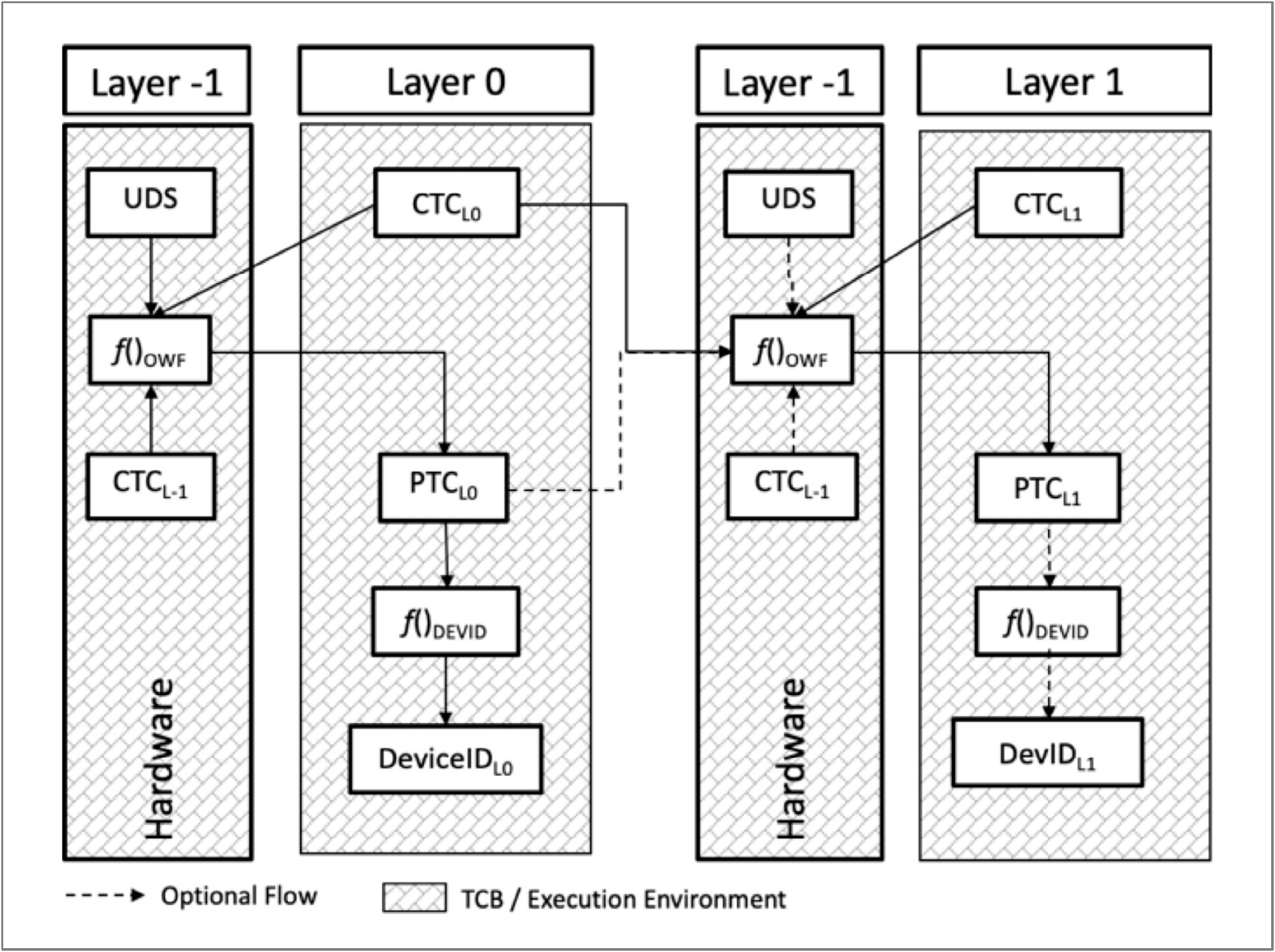}
\caption{Intel SGX layered TCB architecture.}
\label{fig:Intel-DICE-Layering}
\end{figure}

A third example (Figure~\ref{fig:Intel-DICE-Layering}) shows Intel SGX layered TCB architecture. 
In this architecture each layer has access to the Layer -1 TCB. 
The Layer -1 TCB is identified using the CPU product ID and CPU SVN values. 
A UDS provides platform uniqueness. 
The one-way function is always performed by a hardware RoT in Layer -1. 

SGX layers have more flexibility (than the generalized DICE TCB layering) 
in that the PTC computation does not require inclusion of the Layer -1 UDS and CTC. 
However, the Layer -1 TCB supplies a local attestation capability 
that allows any enclave environment the ability to assess layering semantics. 
If layer sequence semantics are important to a user deployment, 
the expected layering can be created. 
Expected layering semantics can be verified using 
a combination of local and remote attestation.

Given a user deployment focus on trustworthy Edge and Cloud computing, 
use of trusted computing techniques that include 
hardware RoT and TCB layering play a vital role. 
Container environments may have widely varied trust properties. 
As Edge and Cloud ecosystems become more democratized and complex, 
reliance on distributed trust becomes essential. 
Attestation is a mechanism whereby verifiers, 
presumable the entity fulfilling an SLA contract 
and the user community they represent, 
seek to manage risk associated with automation complexity and ecosystem diversity.

By incorporating principles of trusted computing 
into a decentralized TCB layer, 
many of the challenges facing application developers seeking 
predictable deployment in Edge and Cloud environments 
can be modularized for ubiquitous availability and 
relied upon for consistent trusted computing behavior.

\section{Use-Case: Gateways for Blockchain Interoperability}
\label{sec:DTCB-usecase}

Given the history of the development of the Internet and of 
computer networks in general (e.g. LANs, WANs),
it is unlikely that the world will settle on one global blockchain system operating universally.
The emerging picture will most likely consist
of ``islands'' of blockchain systems,
which -- like autonomous systems that make-up the Internet -- must be ``stitched'' 
together in some fashion to make a coherent unity.
Similar to packets of messages traversing different paths through the stitched islands of the Internet,
blockchain transactions must be free to traverse the stitched islands of blockchain systems.
This freedom of transactions to traverse or move across blockchain systems is not only
needed to prevent ``asset lock-in'' to a given platform,
but it is crucial from the point of view survivability of these systems as a whole.

Following from the first fundamental goal of the Internet architecture,
the lesson learned there was that {\em interoperability is key to survivability}.
Thus, interoperability is core to the entire value-proposition of blockchain technology.
Interoperability across blockchain systems
must be a requirement -- both at the mechanical level and the value level --
if blockchain systems and technologies are to become
the fundamental infrastructure of the future global commerce (see ~\cite{TradecoinSciAm2018,TradecoinRSOS2018}).

In this section we discuss the use of the DTCB for the purpose
of addressing some of the security challenges pertaining
to {\em blockchain gateways} -- a  notion put forward 
by~\cite{HardjonoLipton2018a} as a counterpart to routing gateways (e.g. BGP Routers) in the Internet.

Blockchain gateways provides a number of potential benefits,
notably in use-cases involving the transferral of assets across different blockchain systems:
\begin{itemize}

\item	{\em Better control over ledger visibility}: 
The first potential benefit of gateways in the context of 
blockchain interoperability is to provide manageable 
control over the visibility (i.e. read access) of 
data residing on the ledger within a given private/permissioned blockchain system. 
The visibility of local ledger data is particularly relevant 
for inter-domain (cross blockchain) transactions, 
in cases where one or both of the blockchain systems 
are private and where the ledger data is confidential.

\item	{\em Trust establishment across distinct blockchain systems}: 
The second potential benefit of gateways in the context of 
blockchain interoperability is to support the establishment of trust 
(i.e. technical-trust) across blockchain autonomous systems. 
This need for inter-domain trust establishment maybe necessary 
for scenarios involving high value transactions. 
Although trust establishment is traditionally 
performed between two devices, there may be use-cases that 
require multiple gateways in one system to simultaneously 
establish group-oriented trust with multiple gateways in another blockchain system.

\item	{\em Peering-points for service contracts}: 
The third potential benefit of gateways in the context of 
blockchain interoperability is to serve as the peering-points that are recognized (called out) 
within peering agreements or contracts. Similar to peering agreements 
between ISPs in Internet routing, new kinds of peering agreements will need to be 
developed for blockchain system interoperability in order that these independent systems 
can interconnect in a secure and reliable fashion.

\end{itemize}

There are several positive benefits of employing the DTCB model for blockchain gateways:
\begin{itemize}

\item	{\em Provide higher assurance to nodes participating in consensus protocols}: 
Augment nodes to employ TCB-related technologies to allow them 
not only to operate in a provable trustworthy manner but 
to allow them to convey this trust in some measurable 
way to external entities (e.g. to other peer nodes and to wallet systems at the end-users).

\item	{\em Provide foundations for TCB-capable nodes to dynamically become Gateways}: 
In order for distinct blockchain systems to interoperate with 
each other to achieve service scale (e.g. for upper layer applications), 
some (or all) nodes in one blockchain system must be able to act as 
gateways to interact with their corresponding gateways in 
a different blockchain system. 

\item	{\em Trust establishment across different blockchain systems using ROT}: 
Employ the properties P1 to P4, DP1 and DP2 (see Section~\ref{sec:PropertiesTechnicalTrust})
to establish trust between two peer gateways, and between two groups of gateways (multi-gateways).

\end{itemize}

\begin{figure}[t]
\centering
\includegraphics[width=0.7\textwidth, trim={0.0cm 0.0cm 0.0cm 0.0cm}, clip]{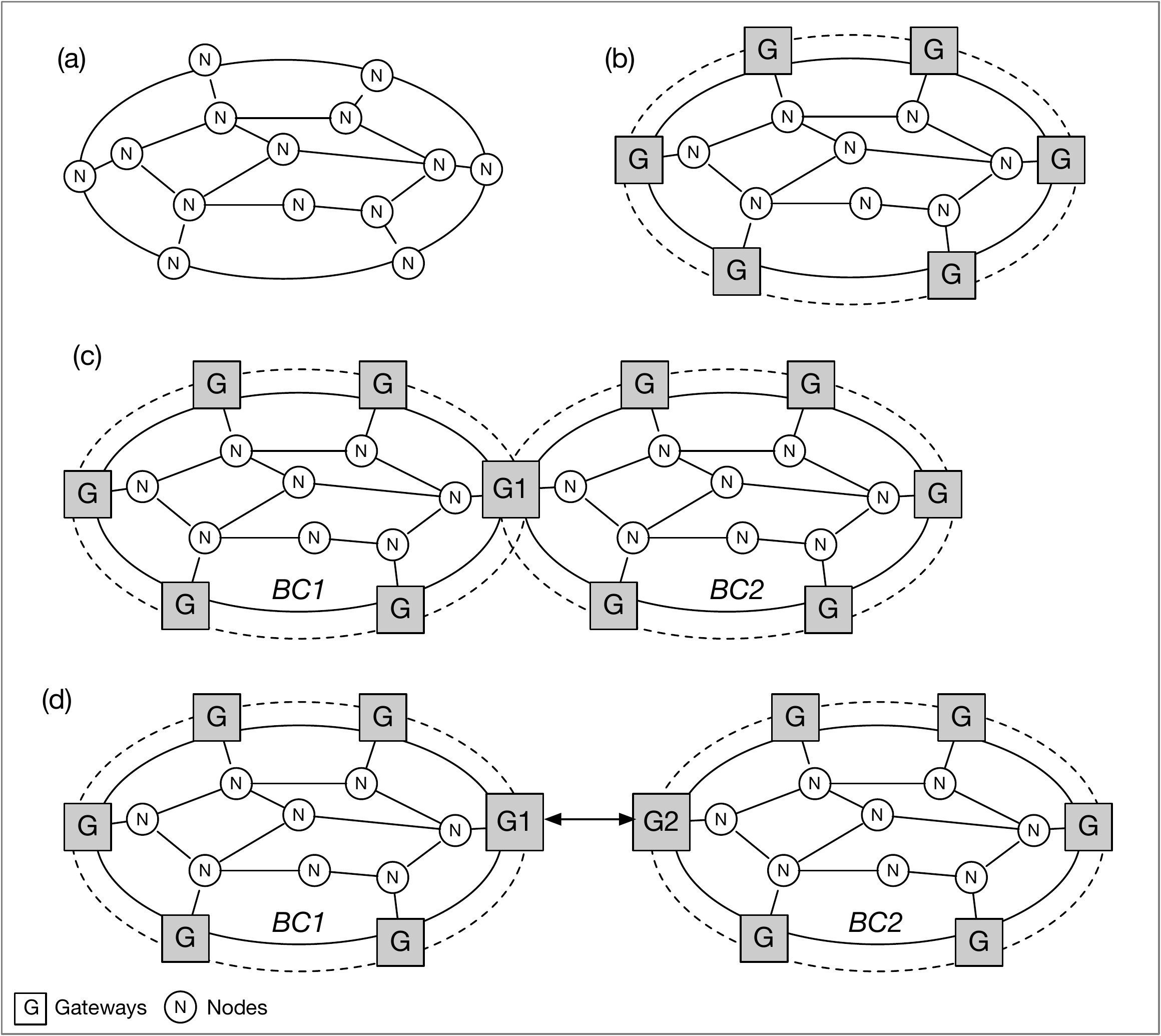}
\caption{Illustration of (a) blockchain nodes in P2P network; (b) Some nodes as Gateways; (c) Node G1 acting as a Bridge; (d) Nodes G1 and G2 as Gateways}
\label{fig:DTCBNodes}
\end{figure}

Similar to the routing autonomous systems,
a blockchain autonomous system may consists of 
numerous nodes that make up the P2P network (Figure~\ref{fig:DTCBNodes}(a)).
Independent of whether the blockchain is permissionless or permissioned,
a number of nodes (or all of the nodes) in the P2P network
must have the capability to handle transactions 
that involve foreign blockchain systems.

Ideally, a node could be a ``member'' of two (or more) blockchain autonomous systems.
In this case, the node would act as a ``bridge'' between the two blockchain autonomous systems (see Figure~\ref{fig:DTCBNodes}(c)).
Although this bridge model has its own security challenges,
in general we cannot assume that such a node will exist for all blockchain autonomous systems configurations.

As such, we assume here that the most realistic deployment configuration 
-- taking into account private/permissioned blockchains --
is for two (2) gateways to interact, where each gateway represents its ``home'' blockchain system.
This is akin to two BGP routers in the Internet belonging to two ISPs respectively, 
where the BGP routers are peered in order for them to exchange route advertisements.
Figure~\ref{fig:DTCBNodes}(d) illustrates the situation in which two nodes G1 and G2 are acting
as gateways between two corresponding blockchain autonomous systems BC1 and BC2.

\subsection{Blockchain Autonomous Systems}

Similar to a routing autonomous system (routing AS)
being composed of one or more routing domains,
we propose viewing a blockchain system as an autonomous system (blockchain AS).

Thus, just as routers in a routing-domain operate one or more routing protocols
to achieve best routes through that domain,
{\em nodes} in a blockchain AS contribute 
to maintaining a shared ledger by running one or more
{\em ledger management protocols} (e.g. consensus algorithms, membership management)
to achieve stability and fast convergence (i.e. confirmation throughput)
of the ledger in that AS.
The division also maps readily into permissioned and permissionless/public blockchains,
where each type of community (and each instance) could be viewed as a separate blockchain AS.

Nodes could therefore be classified from the perspective
of ledger management as operating 
either {\em intra-domain} or {\em inter-domain} (across autonomous systems):
\begin{itemize}

\item	{\em Intra-domain nodes}:
These are nodes and other entities whose main task is maintaining
ledger information and conducting transactions within one blockchain AS.
Examples includes nodes which participate in consensus computations 
(e.g. full mining nodes in Bitcoin),
nodes that ``orchestrate'' consensus computations 
(e.g. Orderers and Endorsers in Hyperledger Fabric, see~\cite{AndroulakiBarger2018}),
and nodes which perform validations only 
(e.g. Validators in Ripple, see~\cite{SchwartzYoungs2014}).

\item	{\em Inter-domain gateways}:
These are nodes and other entities whose main task
is dealing with with transactions
involving different blockchain autonomous systems.
We refer to these nodes as {\em inter-domain gateways} (or simply ``gateways'').

Nodes which are gateways must implement very stringent security requirements
because they interact with other gateways
belonging to different blockchain autonomous systems, 
and thus different administrative jurisdictions,
and potentially different legal jurisdictions.

\end{itemize}

\subsection{Gateways Between Blockchain Systems}

\begin{figure}[t]
\centering
\includegraphics[width=0.9\textwidth, trim={0.0cm 0.0cm 0.0cm 0.0cm}, clip]{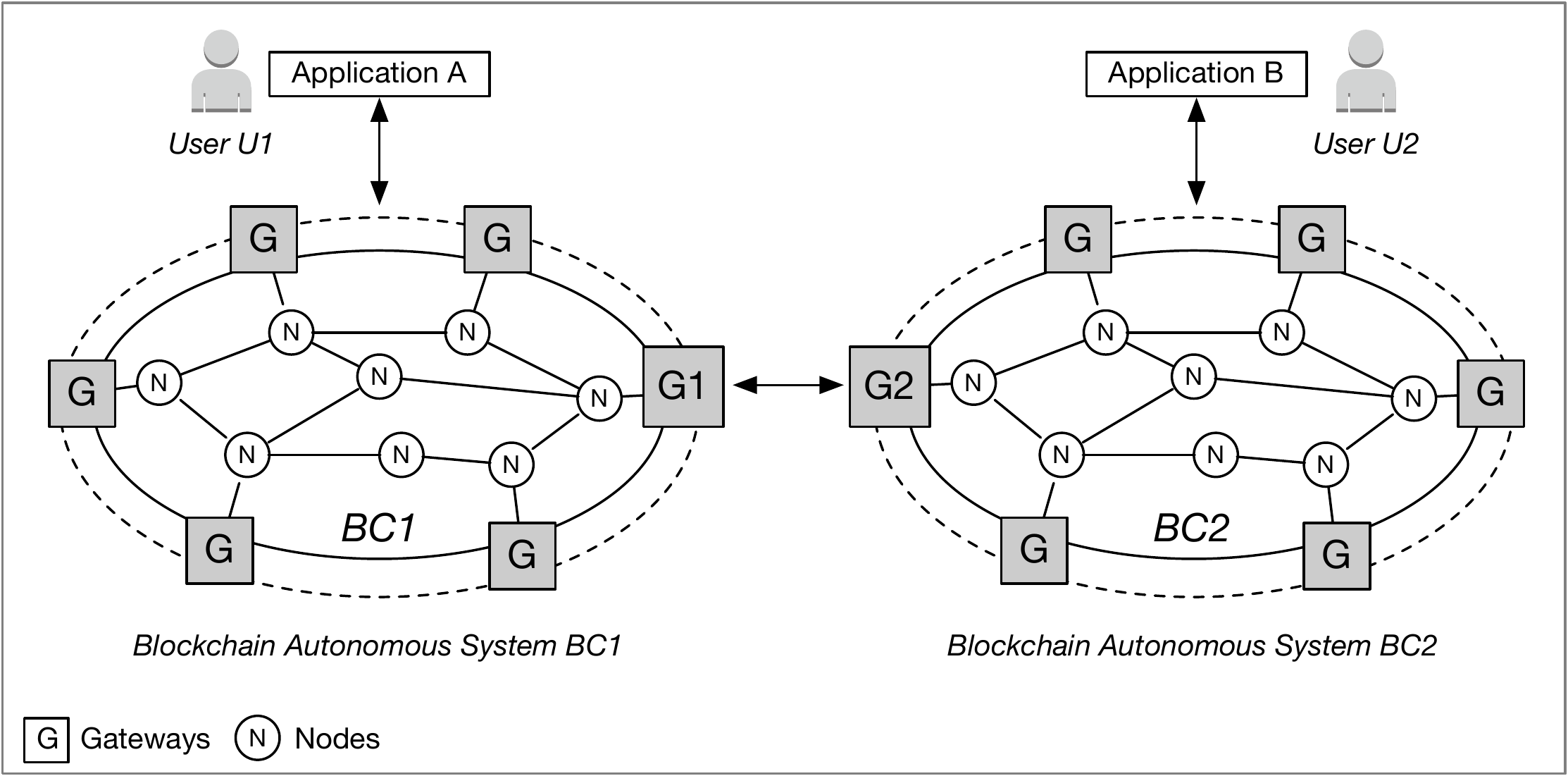}
\caption{Transactions Crossing Blockchain Autonomous Systems}
\label{fig:DTCB-CrossDomain}
\end{figure}

To illustrate,
we sketch a simple example shown in Figure~\ref{fig:DTCB-CrossDomain}
in which an asset recorded in the shared ledger in BC1
is to be transferred to blockchain system BC2.
In Figure~\ref{fig:DTCB-CrossDomain}, 
a user U1 with Application A
has his or her asset ownership (e.g. land title deed)
recorded on the shared-ledger inside blockchain BC1.
The user U1 wishes to transfer legal ownership
of the asset to a different user U2 running Application B,
and to have the asset recoded authoritatively on the shared-ledger
inside blockchain BC2.
We assume both BC1 and BC2 are private/permissioned systems.

In this scenario, the set of gateways in blockchain system BC1 
have agreed to allow G1 to ``speak on behalf'' of BC1.
That is, they have delegated authority to a single gateway G1.
Similarly, G2 has been delegated authority to speak on behalf of blockchain system BC2.

The sketch is as follows (Figure~\ref{fig:DTCB-CrossDomain}):
\begin{enumerate}

\item	The user U1 of Application A initiates the transfer to user U2 running Application B.
This consists of the user U1 transmitting a new (inter-domain) transaction within blockchain system BC1,
addressed to user U2 whose identity (public-key) is present in BC2.
Since the destination of this transaction is a public-key located
outside BC1 (e.g. $BC2$/$PubKey_{U2}$), 
this transaction can only be processed by nodes in BC1 that have
the capability of being a gateway (namely G1).

\item	Gateway G1 notices the pending (unprocessed) transaction destined for a foreign blockchain BC2.
Gateway G1 begins trust establishment with gateway G2 in blockchain system BC2.
Because blockchain system BC1 is a private/permissioned system,
data in its ledger is not visible from outside the blockchain. 
As such, gateway G1 has to create a new public transaction-identifier TxID1 for the asset
that masks the original transaction-identifier recorded on the private shared ledger in BC1.
That is, G1 has to ``mask'' the original transaction-identifier.

\item	After pairwise technical-trust has been established between gateways G1 and G2,
the gateway G2 proceeds to introduce a new registration-transaction for the asset 
into the local ledger of blockchain systems BC2.
This act is essentially ``registering'' the soon-to-arrive asset in the shared ledger of BC2.
This asset is ``locked'' (e.g. using 2-Phase Commit or similar construct) by G2
until G1 settles the transaction in BC1.
While this asset is in locked status in BC2, 
user U2 is prevented from making use of the asset.

\item	After a confirmation has been achieved in the ledger of BC2,
the gateway G2 performs two things.
First, it creates new public transaction-identifier TxID2 that masks
the private transaction-identifier of the asset in BC2;
(ii) Secondly, G2 issues a signed assertion to G1 to the effect that
the asset labelled TxID2 has been recognized and temporarily registered 
in the ledger of BC2.

\item	Upon receiving the signed assertion from G2,
the gateway G1 proceeds to introduce an ``invalidation'' transaction into the ledger of BC1.
In effect, this invalidation transaction marks in the ledger in BC1
that the asset no longer resides in BC1.
The invalidation transaction in BC1 records both TxID1 and TxID2
for future redirections.
This allows for future queriers looking for TxID1 to be redirected to BC2 to obtain TxID2.

\item	After the invalidation-transaction is confirmed on the ledger of BC1,
the gateway G1 issues a signed assertion (to G2) stating 
that local invalidation-transaction
has been confirmed on BC1.
The signed assertion uses the TxID1 and TxID2 identifiers,
which are public transaction identifiers (not the private identifiers inside BC1 and BC2 respectively).

\item	Upon receiving the signed assertion from G1,
the gateway G2 releases the lock on the asset in the ledger of BC1,
thereby allowing its new owner U2 to use the asset.

\end{enumerate}

Several variations of the above example can be devised,
improving the efficiency of the messages and transaction throughput.
The goal of this example, however,
is to illustrate 
(i) the crucial role that gateways G1 and G2 play
in cross-AS transactions;
(ii) the relevance of the DTCB model in securing G1 and G2;
and
(iii) the potential use of the DTCB model in
supporting group-oriented computations
that makes use of several gateways in BC1 simultaneously
for increased resiliency (in contrast to the example using one gateway G1 only).

\subsection{Applications of Features of the DTCB for Gateways and Multi-Gateways}

\begin{figure}[t]
\centering
\includegraphics[width=0.8\textwidth, trim={0.0cm 0.0cm 0.0cm 0.0cm}, clip]{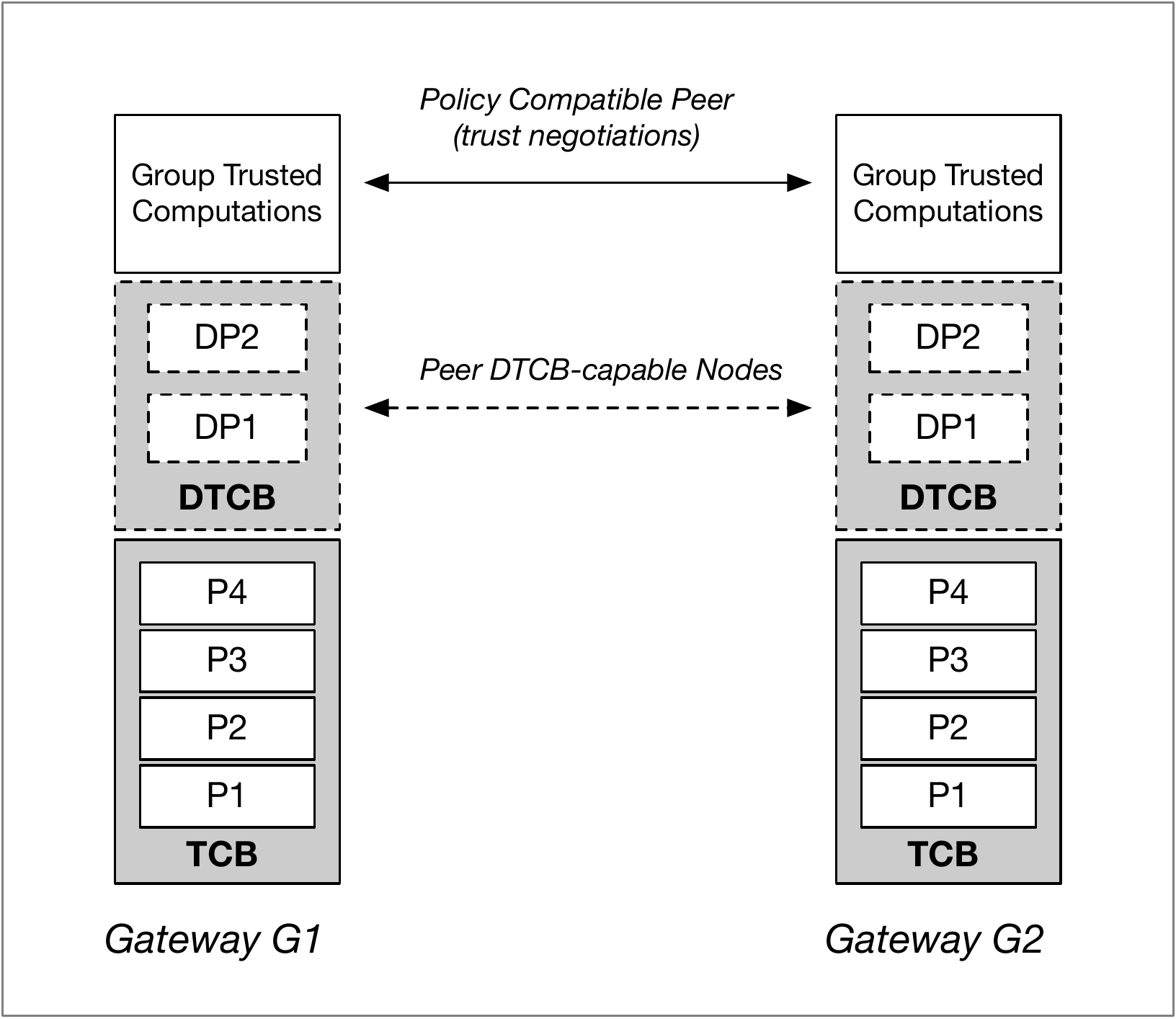}
\caption{Group-Oriented Features of the DTCB for Gateways}
\label{fig:DTCB-Gateways}
\end{figure}

In Section~\ref{sec:PropertiesTechnicalTrust} we discussed a number of properties
that support the decentralized TCB model.
In this section we briefly review the application of those properties to the
blockchain gateways use-case.

As mentioned above,
gateways in Internet routing play an important role for connecting two or more routing autonomous systems,
allowing both route-advertisements and datagrams to flow through the network.
Blockchain gateways, however, have the additional task of acting on behalf
of its blockchain autonomous system in cases of value-carrying transaction that are inter-domain.
This notion of ``acting on behalf'' of a blockchain autonomous system 
introduces several interesting challenges in
both permissionless and permissioned blockchains (Figure~\ref{fig:DTCB-Gateways}):
\begin{itemize}

\item	{\em Proving DTCB properties by a Gateway locally}:
In scenarios where the P2P network of nodes consists
of a mix of DTCB-capable nodes and non-DTCB nodes,
the ability to make use of truthful attestations (Property DP2)
allows DTCB-capable nodes to be distinguished (from non-DTCB nodes) as potential gateways.

\item	{\em Dynamic establishment of Multi-Gateways}:
A given blockchain autonomous system may operate on the basis of the
identification and selection of DTCB-capable nodes as a group gateways or {\em multi-gateways}.
A multi-gateway is a group of DTCB-capable gateways that must act at a unit.
All actions to be taken by any member of the group must be based
on some group-computation.

\item	{\em Matching DTCB properties between gateways in different blockchain autonomous systems}:
As part of technical-trust negotiation and establishment
between gateways G1 in BC1 and gateway G2 in BC2 (see Figure~\ref{fig:DTCB-CrossDomain}),
G1 and G2 must validate each other's DTCB properties.
This may include mutually validating all of the roots of trust in properties P1, P2 and P3
in each gateway.

\item	{\em Proving execution of consensus algorithm under shielded processing}:
The property P2 (see Section~\ref{sec:PropertiesTechnicalTrust})
allows DTCB-capable nodes to (i) safely execute consensus algorithms
(and other related computations), and 
(ii) to prove that these algorithms were executed under shielding.

\item	{\em Majority gateway consensus for sensitive transactions}:
The set of DTCB-capable gateways may collectively implement governance rules
that require sensitive inter-domain transactions 
to be ``approved by'' the majority of these gateways.
That is, gateways may become a subcommunity of nodes,
whose majority quorum is needed for inter-domain transactions.

\item	{\em Gateway group-signatures}:
The subcommunity DTCB-capable gateways may collectively
implement one or more {\em group-oriented digital signatures}
(for example, see~\cite{Desmedt1987}).
This feature maybe useful in high-value inter-domain transactions
which require high assurance.

\item	{\em Membership-verifiability with gateway anonymity}:
Depending on the specific implementation of 
properties P3 and DP1 (see Section~\ref{sec:PropertiesTechnicalTrust}),
a gateway can prove (a) that it is a DTCB-capable nodes and (b) that it belong to the gateways subcommunity
without revealing its identity.
This feature maybe useful in the first stages of 
gateway-to-gateway trust negotiations,
where an honest gateway is not able to know if its
opposite gateway is truly DTCB-capable.
This process may also be used to reduce DDOS attacks on gateways,
in which rogue entities (i.e. fake non-DTCB machines)
exhaust the resources of legitimate gateways through opening fake handshakes.

\item	{\em DTCB manifest as condition for peering agreements}:
For peering between two permissioned/private blockchain autonomous systems,
a manifest of the list of minimal hardwares and softwares for DTCB-capable nodes
provides a way for organizations to build Service Level Agreements (SLAs)
based on measurable technical-trust.
Property DP2 ensures that a DTCB-capable node can truthfully report
its manifest to an external entity as part of satisfying the peering agreement.

\item	{\em DTCB-assisted multi-party computation}:
A set of DTCB-capable gateways can collectively use properties P1, P2 and DP2
to jointly perform a given multi-party computation (MPC) (see~\cite{Lindell2003}).
For example, a given MPC computation maybe designed to yield a common cryptographic key $K_{BC1,BC2}$
shared between two opposing groups of gateways
(e.g. one group in BC1 and the other in BC2).
This would allow any gateway in BC1 holding key $K_{BC1,BC2}$
to begin interacting with any gateway in BC2 holding the same key.

\end{itemize}


\section{Conclusion and Further Considerations}

Although there has been significant interest and media attention placed on the area of blockchain technology,
there remains a number open issues that needs to be addressed.
These range from the problem of the concentration of hash-power,
anonymity of entities,
lack of good key management,
to the lack of measurable technical-trust required for business agreements and SLAs.
As such, there is a strong need for these issues to be addressed
before the blockchain technology can be the basis for the future
global financial infrastructure.

We believe that a decentralized TCB model is the appropriate technological foundation
for providing technical-trust for the blockchain infrastructure.
This includes the hardening of individual nodes and systems in the blockchain infrastructure,
to providing support for secure group-oriented computations -- including consensus algorithms
and multi-party computations  --
for nodes that make-up a blockchain system.
This paper devoted considerable attention to the virtualized cloud environments
because it is likely that much of the future blockchain infrastructure
may operate in cloud environments.

Finally, we discussed the role of gateways in blockchain autonomous systems
as the modern counterpart of routing autonomous systems.
Aside from providing controlled visibility over data recorded in the shared ledger
of private/permissioned blockchains,
gateways are needed for the interoperability of independent blockchain systems.
We believe that the decentralized TCB model provides the basis for developing solutions
that support gateways in establishing technical-trust with each other.
The ability to express security quality in some measure based on the DTCB
allows blockchain infrastructure owners and operators
to develop a common legal framework for establishing peering connectivity, and therefore scale.
This is how the Internet and IP routing evolved over the past three decades,
and we believe this is how the blockchain infrastructure will also evolve.

\section*{Acknowledgments}
We thank Prof Sandy Pentland and Prof Alexander Lipton at MIT for support and insights into this work.
We thank the following for ideas, critiques and support: 
Anne Kim, Justin Anderson, Michael Casey, Sandro Lera (MIT);
Jerry Cuomo, Gari Singh, Mac Devine, Jeb Linton (IBM);
Jan Camenisch (Dfinity).
We also thank the numerous colleagues at Intel who have made 
valuable contributions to trusted computing including 
Frank McKeen, Claire Vishik, David Grawrock, Vinnie Scarlata, 
Simon Johnson, David M. Wheeler and Abhilasha Bhargav-Spantzel 
as well as those who have helped develop our understanding of 
cloud, edge and IoT computing including Francesc Guim, 
Dario Sabella, Mikko Ylinen and Kapil Sood. 
We especially want to thank 
Jennifer E. Koerv for being an advocate and champion of this work.



\end{document}